\documentclass[runningheads]{llncs}

\usepackage{graphicx}
\usepackage[dvipsnames]{xcolor}
\usepackage{svg}
\usepackage{amsmath}
\usepackage{amssymb}
\usepackage{subfig}
\usepackage{booktabs}
\usepackage[utf8]{inputenc}
\usepackage[T1]{fontenc}
\usepackage[ruled,vlined,linesnumbered,noresetcount]{algorithm2e}
\usepackage{multirow}

\DeclareMathOperator{\argmax}{arg\,max}

\newcommand{\orcid}[1]{\href{https://orcid.org/#1}{\textcolor[HTML]{A6CE39}{\includegraphics[width=10pt]{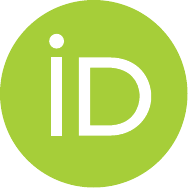}}}}
\usepackage{hyperref}

\begin{document}
\newcommand*\samethanks[1][\value{footnote}]{\footnotemark[#1]}
\title{CTC-Segmentation of Large Corpora for German End-to-end Speech Recognition}
\titlerunning{CTC-Segmentation of Large Corpora for German Speech Recognition}
%
\author{ Ludwig K{\"u}rzinger\orcid{0000-0001-5312-3870}\thanks{These authors contributed equally to this work.} \and
	Dominik Winkelbauer\samethanks \and
	Lujun Li\orcid{0000-0002-0641-3178} \and
Tobias Watzel\orcid{0000-0002-3552-3325} \and
Gerhard Rigoll\orcid{0000-0003-1096-1596}}
\authorrunning{L. K{\"u}rzinger et al.}
%
\institute{Institute for Human-Machine Communication,\\ Technische Universit{\"a}t M{\"u}nchen, Germany\\
\email{\{ludwig.kuerzinger,dominik.winkelbauer\}@tum.de}}

\maketitle
\begin{abstract}
Recent end-to-end Automatic Speech Recognition (ASR) systems demonstrated the ability to outperform conventional hybrid DNN/ HMM ASR.
Aside from architectural improvements in those systems, those models grew in terms of depth, parameters and model capacity.
However, these models also require more training data to achieve comparable performance.

In this work, we combine freely available corpora for German speech recognition, including yet unlabeled speech data, to a big dataset of over $1700$h of speech data.
For data preparation, we propose a two-stage approach that uses an ASR model pre-trained with Connectionist Temporal Classification (CTC) to boot-strap more training data from unsegmented or unlabeled training data.
Utterances are then extracted from label probabilities obtained from the network trained with CTC to determine segment alignments.
With this training data, we trained a hybrid CTC/attention Transformer model that achieves $12.8\%$ WER on the Tuda-DE test set,
surpassing the previous baseline of $14.4\%$ of conventional hybrid DNN/HMM ASR.%
\footnote{This is a preprint article. The full paper~\cite{ctcsegmentation} can be found at \\ {\url{https://doi.org/10.1007/978-3-030-60276-5_27}}}
\end{abstract}

\section{Introduction}
Conventional speech recognition systems combine Deep Neural Networks (DNN) with Hidden Markov Models (HMM).
The DNN serves as an acoustic model that infers classes, or their posterior probabilities respectively, originating from handcrafted HMMs and complex linguistic models.
Hybrid DNN/HMM models also require multiple processing steps during training to refine frame-wise acoustic model labels.
In comparison to hybrid DNN/HMM systems, end-to-end ASR simplifies training and decoding by directly inferring sequences of letters, or tokens, given a speech signal.
For training, end-to-end systems only require the raw text corresponding to an utterance.
Connectionist Temporal Classification (CTC) is a popular loss function to train end-to-end ASR architectures~\cite{Graves2006}.
In principle, its concept is similar to a HMM, the label sequence is modeled as sequence of states, and during training, a slightly modified forward-backward algorithm is used in the calculation of CTC loss.
Another popular approach for end-to-end ASR is to directly infer letter sequences, as employed in attention-based encoder-decoder architectures~\cite{Chan2016}.
Hybrid CTC/attention ASR architectures combine these two approaches~\cite{Watanabe2017}.

End-to-end models also require more training data to learn acoustic representations.
Many large corpora, such as Librispeech or TEDlium, are provided as large audio files partitioned into segments that contain speech with transcriptions.
Although end-to-end systems do not need frame-wise temporal alignment or segmentation, an utterance-wise alignment between audio and text is necessary.
To reduce training complexity, previous works used frameworks like sphinx~\cite{Lamere2003} or MAUS~\cite{schiel1999} to partition speech data into sentence-length segments, each containing an utterance.
Those frameworks determine the start and the end of a sentence from acoustic models (often HMMs) and the Viterbi algorithm.
However, there are three disadvantages in using these for end-to-end ASR:
(1) As only words in the lexicon can be detected, the segmentation tool needs a strategy for out-of-vocabulary words.
(2) Scaling the Viterbi algorithm to generate alignments within larger audio files requires additional mitigations.
(3) As these algorithms provide \emph{forced} alignments, they assume that the audio contains only the text which should be aligned; but for most public domain audio this is not the case.
So do for example all audio files from the Librivox dataset contain an additional prologue and epilogue where the speaker lists his name, the book title and the license.
It might also be the case that the speaker skips some sentences or adds new ones due to different text versions.
Therefore, aligning segments of large datasets, such as TEDlium~\cite{Rousseau2014}, is done in multiple iterations that often include manual examination.
Unfortunately, this process is tedious and error prone;
{for example, by inspection of the SWC corpus, some of those automatically generated transcriptions are missing words in the transcription.}

We aim for a method to extract labeled utterances in the form of correctly aligned segments from large audio files.
To achieve this, we propose CTC-segmentation, an algorithm to correctly align start and end of utterance segments, supported by a CTC-based end-to-end ASR network
\footnote{The source code underlying this work is available at~\url{https://github.com/cornerfarmer/ctc_segmentation}}.
Furthermore, we demonstrate additional data cleanup steps for German language orthography.
\textbf{Our contributions are:}
\begin{itemize}
    \item We propose CTC-segmentation, a \emph{scalable} method to extract utterance segments from speech corpora.
    In comparison to other automated segmentation tools, alignments generated with CTC-segmentation were observed to more closely correspond to manually segmented utterances.
    \item We extended and refined the existing recipe from the ASR toolkit kaldi with a collection of open source German corpora by two additional corpora, namely \emph{Librivox} and \emph{CommonVoice},
    and ported it to the end-to-end ASR toolkit ESPnet.
\end{itemize}

\section{Related Work}

\subsection{Speech Recognition for German}
Milde et al.~\cite{Milde2018} proposed to combine freely available German language speech corpora into an \emph{open source} German speech recognition system.
A more detailed description of the German datasets can be found in~\cite{Milde2018}, of which we give a short summary:
\begin{itemize}
    \item The Tuda-DE dataset~\cite{RadeckArneth2015} combines recordings of multiple sentences concerning various topics spoken by 180 speakers using five microphones.
    \item The Spoken Wikipedia Corpus (SWC,~\cite{Baumann2016}) is an open source summary of recordings of different Wikipedia articles made by volunteers.
    The transcription already includes alignment notations between audio and text, but as these alignments were often incorrect, Milde et al. re-aligned utterance segments using the Sphinx speech recognizer~\cite{Lamere2003}.
    \item The M-AILABS Speech Dataset~\cite{Solak2019} {mostly} consists of utterances extracted from political speeches and audio books from Librivox.
    Audio and text has been aligned by using synthetically generated audio (TTS) based on the text and by manually removing intro and outro.
\end{itemize}
In this work, we additionally combine the following German speech corpora:
\begin{itemize}
    \item CommonVoice dataset~\cite{Ardila2019} consists of utterances recorded and verified by volunteers;
    therefore, an utterance-wise alignment already exists.
    \item Librivox~\cite{librivox} is a platform for volunteers to publish their recordings of reading public domain books.
    All recordings are published under a Creative Common license.
    We use audio recordings of $579$ books.
    The corresponding texts are retrieved from Project Gutenberg-DE~\cite{Gutenberg2019} that hosts a database of books in the public domain.
\end{itemize}
Milde et al.~\cite{Milde2018} mainly used a conventional DNN/HMM model, as provided by the kaldi toolkit~\cite{Povey2011}.
Denisov et al.~\cite{Denisov2019} used a similar collection of German language corpora that additionally includes non-free pre-labeled speech corpora.
Their ASR tool \emph{IMS Speech} is based on a hybrid CTC/attention ASR architecture using the BLSTM model with location-aware attention as proposed by Watanabe et al.~\cite{Watanabe2017}.
The architecture used in our work also is based on the hybrid CTC/attention ASR of the ESPnet toolkit~\cite{Watanabe2018},
however, in combination with the Transformer architecture~\cite{Vaswani2017} that uses self-attention.
As we only give a short description of its architecture, an in-detail description of the Transformer model is given by Karita et al.~\cite{Karita2019}.

\subsection{Alignment and Segmentation Methods}
There are several tools to extract labeled utterance segments from speech corpora.
The Munich Automatic Segmentation (MAUS) system \cite{schiel1999} first transforms the given transcript into a graph representing different sequences of phones by applying predefined rules. 
Afterwards, the actual alignment is estimated by finding the most probable path using a set of HMMs and pretrained acoustic models.
Gentle works in a similar way, but while MAUS uses HTK \cite{young1993htk}, Gentle is built on top of Kaldi \cite{Povey2011}.
Both methods yield phone-wise alignments.
Aeneas \cite{aeneas2017} uses a different approach:
It first converts the given transcript into audio by using text-to-speech (TTS) and then uses the Dynamic Time Warping (DTW) algorithm to align the synthetic and the actual audio by warping the time axis. 
In this way it is possible to estimate begin and end of given utterance within the audio file.

We propose to use a CTC-based network for segmentation.
CTC was originally proposed as a loss function to train RNNs on unsegmented data.
At the same time, using CTC as a segmentation algorithm was also proposed by Graves et al.~\cite{Graves2006}.
However, to the best knowledge of the authors, while the CTC algorithm is widely used for end-to-end speech recognition,
there is not yet a segmentation tool for speech audio based on CTC.

\section{Methodology}

\subsection{CTC-Segmentation of Utterances}
\label{sec:alignment}
The following paragraphs describe CTC-segmentation, an algorithm to extract proper audio-text alignments in the presence of additional unknown speech sections at the beginning or end of the audio recording.
It uses a CTC-based end-to-end network that was trained on already aligned data beforehand, e.g., as provided by a CTC/attention ASR system.
For a given audio recording the CTC network generates frame-based character posteriors $p(c | t, x_{1:T})$.
From these probabilities, we compute via dynamic programming all possible maximum joint probabilities $k_{t,j}$ for aligning the text until character index $j\in [1;M]$ to the audio up to frame $t\in [1;T]$.
Probabilities are mapped into a trellis diagram by the following rules:
\begin{align}
k_{t,j} &= 
\begin{cases}
\max(k_{t - 1,j} \cdot p( blank | t), \, k_{t - 1,j - 1} \cdot p(c_j | t)) & \raisebox{0pt}{$\text{if } t > 0 \wedge j > 0$} \\
0 & \text{if } t = 0 \wedge j > 0 \\
1 & \text{if } j = 0 
\end{cases}
\end{align}
The maximum joint probability at a point is computed by taking the most probable of the two possible transitions:
Either only a blank symbol or the next character is consumed.
The transition cost for staying at the first character is set to zero, to align the transcription start to an arbitrary point of the audio file.

The character-wise alignment is then calculated by backtracking, starting off the most probable temporal position of the last character in the transcription, i.e, $t=\argmax_{t'} k_{t',M}$.
Transitions with the highest probability then determine the alignment $a_t$ of the audio frame $t$ to its corresponding character from the text, such that
\begin{equation}
a_{t} = \begin{cases}
M - 1 & \text{if } t \geqslant \argmax_{t'}(k_{t',M-1})\\
a_{t+1} & \text{if } k_{t,a_{t+1}} \cdot p( blank | t + 1) > k_{t,a_{t+1} - 1} \cdot p(c_j | t + 1)\\
a_{t+1} - 1 & \text{else}
\end{cases}.
\end{equation}
As this algorithm yields a probability  $\rho_t$ for every audio frame being aligned in a given way,
a \emph{confidence score} $s_{\text{seg}}$ for each segment is derived to sort out utterances with deviations between speech and corresponding text, that is calculated as
\begin{equation}
s_{\text{seg}} = \min_j m_j \quad \text{with}\quad m_j = \frac{1}{L} \sum_{t=jL}^{(j+1)L}{\rho_t}.
\end{equation}
Here, audio frames that were segmented to correspond to a given utterance are first split into parts of length $L$.
For each of these parts, a mean value $m_j$ based on the frame-wise probabilities $\rho_t$ is calculated.
The total probability $s_{\text{seg}}$ for a given utterance is defined as the minimum of these probabilities per part $m_j$.
This method inflicts a penalty on the confidence score on mismatch, e.g., even if a single word is missing in the transcription of a long utterance.

The complexity of the alignment algorithm is reduced from $O(M \cdot N)$ to $O(M)$ by using the heuristic that the ratio between the aligned audio and text position is nearly constant.
Instead of calculating all probabilities $k_{t,j}$, for every character position $j$ one only considers the audio frames in the interval $[t - W/2, t + W/2]$ with $t = jN/M$ as the audio position proportional to a given character position and the window size $W$.

\subsection{Data cleaning and text preparation}
The ground truth text from free corpora, such as Librivox or the SWC corpus, is often not directly usable for ASR and has therefore to be cleaned. To maximize generalization to the Tuda-DE test dataset, this is done in a way to match the style of the ground truth text used in Tuda-DE, which only consists of letters, i.e. a-z and umlauts ({\"a}, {\"u}, {\"o}, {\ss}).
Punctuation characters are removed and all sentences with different letters are taken out of the dataset. 
All abbreviations and units are replaced with their full spoken equivalent.
Furthermore, all numbers are replaced by their full spoken equivalent.
Here it is also necessary to consider different cases, as this might influence the suffix of the resulting word. 
Say, \emph{``\textbf{18}00 Soldaten''} needs to be replaced by \emph{``\textbf{eintausendacht}hundert Soldaten''}, whereas \emph{``Es war \textbf{18}00''} is replaced according to its pronunciation by \emph{``Es war \textbf{achtzehn}hundert''}.
The correct case can be determined from neighboring words with simple heuristics.
For this, the NLP tagger provided by the spacy framework [7] is used.

Another issue arised due to old German orthography.
Text obtained from Librivox is due to its expired copyright usually at least 70 years old and uses old German spelling rules.
For an automated transition to the reformed German orthography,
we implemented a self-updating lookup-table of letter replacements.
This list was compiled based on a list of known German words from correctly spelled text.

\section{Evaluation and Results}

\subsection{Alignment evaluation}

In this section, we evaluate how well the proposed CTC-segmentation algorithm aligns utterance-wise text and audio.
Evaluation is done on the dev and test set of the TEDlium v2 dataset~\cite{Rousseau2014}, that consist of recordings from 19 unique speakers that talk in front of an audience.
This corpus contains labeled sentence-length utterances, each with the information of start and end of its segment in the audio recording.
As these alignments have been done manually, we use them as reference for the evaluation of the forced alignment algorithms.
The comparison is done based on three parameters:
the mean deviation of the predicted start or end from ground truth, its standard deviation and the ratio of predictions which are at maximum 0.5 seconds apart from ground truth.
To evaluate the impact of the ASR model on CTC-segmentation, we include both BLSTM as well as Transformer models in the comparison.
The pre-trained models%
\footnote{Configuration of the pre-trained models:
The Transformer model has a self-attention encoder with $12$ layers of each $2048$ units.
The BLSTM model has a BLSTMP encoder containing $4$ layers with each $1024$ units, with sub-sampling in the second and third layer.}
were provided by the ESPnet toolkit~\cite{Watanabe2018}.
We compare our approach with three existing forced alignment methods from literature: MAUS, Gentle and Aeneas.
To get utterance-wise from phone-wise alignments, we determine the begin time of the first phone and the end time of the last phone of the given utterance.
As can be seen in Tab.~\ref{table:align_results}, segment alignments generated by CTC-segmentation correspond significantly closer to ground truth compared to the segments generated by all other tested alignment algorithms.

\begin{table}[tpb]
	\centering
	\caption{Accuracy of different alignment methods on the dev and test set of TEDlium v2,
	compared via the mean deviation from ground truth, its standard deviation and the ratio of predictions which are at maximum $0.5$ seconds apart from ground truth.}
	\begin{tabular}{l |l |l| l}\toprule
    & \textbf{Mean} & \textbf{Std} & \textbf{<0.5s}  \\ 
    \midrule
    \multicolumn{4}{l}{\textbf{Conventional Segmentation Approaches}} \\
    \midrule
	MAUS (HMM-based using HTK) & 1.38s & 11.62 & 74.1\% \\
	Aeneas (DTW-based) & 9.01s & 38.47 & 64.7\% \\
	Gentle (HMM-based using kaldi) & 0.41s & 1.97 & 82.0\% \\ 
	\midrule
	\multicolumn{4}{l}{\textbf{CTC-Segmentation (Ours)}} \\
    \midrule
	Hybrid CTC/att. BLSTM trained on TEDlium v2 & 0.34s & 1.16 & 90.1\% \\
	Hybrid CTC/att. Transformer trained on TEDlium v2 & 0.31s & 0.85 & 88.8\% \\
	Hybrid CTC/att. Transformer trained on Librispeech & 0.35s & 0.68 & 85.1\% \\
	\bottomrule 
    \end{tabular}
	\label{table:align_results}
\end{table}

\begin{figure}[tbp]
  \centering
  \includegraphics[width=.65\textwidth]{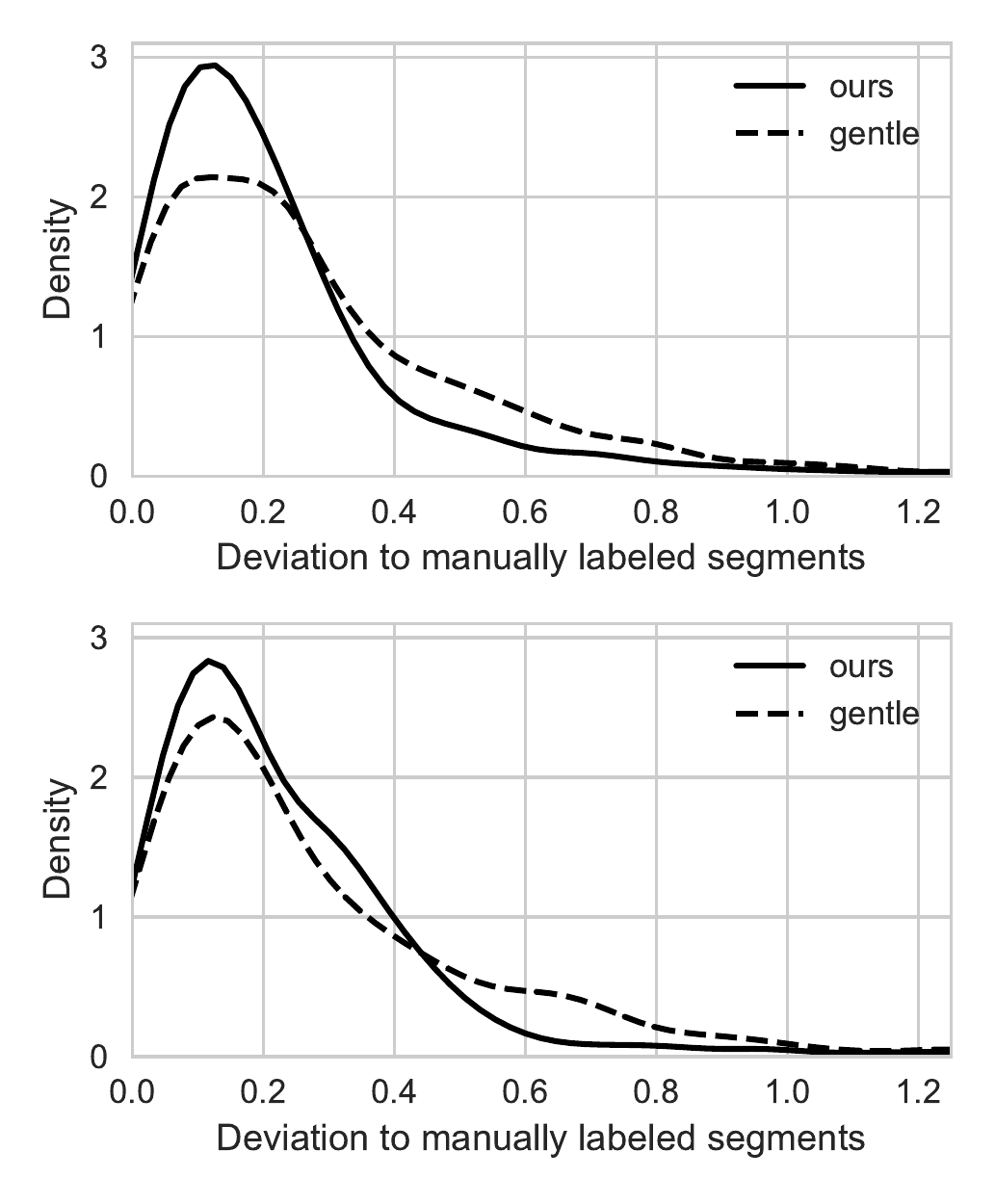}
  \caption{Relative deviation, denoted in seconds, of segments generated by Gentle and our CTC-segmentation compared to manually labeled segments from TEDlium 2.
  CTC-segmentation exhibited a greater accuracy to the start of the segment (top) in comparison with Gentle;
  an also was observed to be slightly more accurate towards the end of the segments (bottom).
  The $y$ axis denotes density in a histogram with $60$ bins.}
  \label{fig:alignment_density}
\end{figure}

Fig.~\ref{fig:alignment_density} visualizes the density of segmentation timing deviations across all predictions.
We thereby compare our approach using the LSTM-based model trained on TEDlium v2 with the Gentle alignment tool.
It can be seen that both approaches have timing deviations smaller than one second for most predictions.
Apart from that, our approach has a higher density in deviations between 0 and 0.5 seconds, while it is the other way around in the interval from 0.5 to 1 second.
This indicates that our approach generates more accurately aligned segments when compared to Viterbi- or DTW-based algorithms.

As explained in section \ref{sec:alignment}, one of the main motivations for CTC-segmentation is to determine utterance segments in a robust manner, regardless of preambles or deviating transcriptions.
To simulate such cases using the TEDlium v2 dev and test set,
we prepended the last $N$ seconds of every audio file before its start and appended the first $M$ seconds to its end.
Hereby, $N$ and $M$ are randomly sampled from the interval $[10, 30]s$.
Table \ref{table:align_results_augm} shows how the same algorithms perform on this altered dataset.
Especially the accuracy of the alignment tools MAUS and Aeneas drops drastically when additional unknown parts of the audio recording are added.
Gentle and our method however are able to retain their alignment abilities in such cases.

To conclude both experiments, alignments generated by CTC-segmentation correspond closer to the ground truth compared to DTW and HMM based methods,
independent of the used architecture and training set.
By inspection, 
the quality of obtained alignments varies slightly across domains and conditions:
The Transformer model with a more powerful encoder performs better compared to the BLSTM model.
Also, the alignments of a model trained on the TEDlium v2 corpus are more accurate on average on its corresponding test and dev set;
this corpus contains more reverberation and noise from an audience than the Librispeech corpus.

\begin{table}[tb]
	\centering
	\caption{Different alignment methods on the augmented dev and test set of TEDlium v2.
	Similar to the evaluation procedure as in Tab.~\ref{table:align_results},
	but the audio samples are augmented by adding random speech parts to their start and end.
	In this the robustness of the different approaches is evaluated.}
	\begin{tabular}{l |l |l| l}\toprule
    & \textbf{Mean} & \textbf{Std} & \textbf{<0.5s}  \\ 
    \midrule
    \multicolumn{4}{l}{\textbf{Existing methods}} \\
    \midrule
	MAUS (HMM-based using HTK) & 3.18s & 18.97 & 66.9 \% \\
	Aeneas (DTW-based) & 10.91s & 40.50 & 62.2 \%\\
	Gentle (HMM-based using kaldi) & 0.46s & 2.40 & 81.7 \%\\ 
	\midrule
	\multicolumn{4}{l}{\textbf{CTC-Segmentation (Ours)}} \\
    \midrule
	BLSTM trained on TEDlium v2 & 0.40s & 1.63 & 89.3 \% \\
	Transformer trained on TEDlium v2 & 0.35s & 1.38 & 89.2 \% \\
	Transformer trained on Librispeech & 0.40s & 1.21 & 84.2 \% \\
	\bottomrule 
    \end{tabular}
\label{table:align_results_augm}
\end{table}

\subsection{Composition of German Corpora for Training}
Model evaluation is performed on multiple combinations of datasets, listed in Tab.\ref{table:datasets}.
Thereby we build upon the corpora collection used by Milde et al.~\cite{Milde2018}, namely, Tuda-DE, SWC and M-AILABS.
As~\cite{Milde2018}, we also neglect recordings made by the Realtek microphone due to bad quality.
Additional to these three corpora, we train our model on Common Voice and Librivox.
Data preparation of the Common Voice dataset only required to post-process the ground truth text by replacing all numbers by their full spoken equivalent.
As the Viterbi-alignment provided by~\cite{Milde2018} for SWC is not perfect, with some utterances missing its first words in the transcription, we realign and clean the data using CTC-segmentation, as in Sec.~\ref{sec:alignment}.
Utterance alignments with a confidence score $s_{\text{seg}}$ lower than $0.22$, corresponding to $-1.5$ in log space, were discarded.
To perform CTC-segmentation on the Librivox corpus,
we combined the audio files with the corresponding ground truth text pieces from Project Gutenberg-DE~\cite{Gutenberg2019}.
Comparable evaluation results were obtained from decoding the Tuda-DE dev and test sets, as also used in~\cite{Milde2018}.

In total, the cumulative size of these corpora spans up to $1772$h, of which we use three partially overlapping subsets for training:
In the first configuration that includes $649$h of speech data, we use the selection as provided by Milde et al. that includes Tuda-DE, SWC and M-AILABS.
The second subset is created by adding the CommonVoice corpus, resulting in $968$h of training data.
The third selection conjoins the Tuda-DE corpus and CommonVoice with the two CTC-segmented corpora, SWC and Librivox, to $1460$h of speech data.

\begin{table}[htb!]%
  \centering
  \caption{Datasets used for training and evaluation.}%
\begin{tabular}{r | c |r|r|r}\toprule
\multicolumn{2}{c|}{\textbf{Datasets}} & \textbf{Length} & \textbf{Speakers} & \textbf{Utterances} \\ \midrule
	Tuda-DE train~\cite{RadeckArneth2015}& TD & $127$h &  $147$ & $55497$ \\
	Tuda-DE dev~\cite{RadeckArneth2015}& dev & $9$h & $16$ & $3678$ \\
	Tuda-DE test~\cite{RadeckArneth2015}& test & $10$h & $17$ &  $4100$ \\
    SWC~\cite{Baumann2016}, aligned by~\cite{Milde2018} & SW & $285$h & $363$ & $171380$ \\   
    M-ailabs~\cite{Solak2019}& MA & $237$h & $29$ & $118521$ \\
    Common Voice~\cite{Ardila2019}& CV & $319$h & $4852$ & $279516$ \\
    CTC-segmented SWC & SW* & $210$h & $363$ & $78214$ \\ 
    CTC-segmented Librivox~\cite{Gutenberg2019,librivox} & LV* & $804$h & $251$ & $368532$ \\ \bottomrule 
\end{tabular}
  \label{table:datasets}
\end{table}

\subsection{ASR configuration}

For all experiments, the hybrid CTC/attention architecture with the Transformer is used.
It consists of a 12 layer encoder and a 6 layer decoder, both with 2048 units in each layer; attention blocks contain 4 heads to each 256 units\footnote{The default configuration of the Transformer model at ESPnet v.0.5.3}.
All models were trained for 23 epochs using the noam optimizer.
We did not use data augmentation, such as SpecAugment.
At inference time, the decoding of the test and dev set is done using beam search with beam size of $16$.
To further improve the results on the test and dev set, a language model was used to guide the beam search.
Language models with two sizes were used in decoding. 
The RNNLM language models were trained on the same text corpus as used in~\cite{Milde2018} for 20 epochs.
The first RNNLM has two layers with $650$ LSTM units per layer. It achieves a perplexity of 8.53.
The second RNNLM consists of four layer of each $1024$ units, with a perplexity of 6.46.

\begin{table}[tb!]
\centering
	\caption{A comparison of using different dataset combinations.
	Word error rates are in percent and evaluated on the Tuda-DE test and dev set.}
\begin{tabular}{llllllr|l|l|c|c}\toprule
\multicolumn{7}{c|}{Datasets} & \multirow{2}{*}{ASR model} & \multirow{2}{*}{LM} & \multicolumn{2}{c}{Tuda-DE} \\
TD & SW & MA & CV & SW* & LV* & h &  &  & dev & test \\
\midrule
\checkmark &   \checkmark  &  -    &  -  &  -  &  -  & $412$ &   TDNN-HMM~\cite{Milde2018} & 4-gram KN      &  15.3 & 16.5 \\
\checkmark &   \checkmark  &  -    &  -  &  -  &  -  & $412$ &   TDNN-HMM~\cite{Milde2018} & LSTM ($2\times 1024$)  &  13.1  & 14.4	 \\
\checkmark &   \checkmark  &  \checkmark    &  -  &  -  &  -  & $649$ &   TDNN-HMM~\cite{Milde2018} & 4-gram KN & 14.8 & 15.9  \\
\midrule
\checkmark &   \checkmark  &  \checkmark    & -  &  -  &  -  & $649$ &   Transformer & RNNLM $(2 \times 650)$ & 16.4 & 17.2	 \\
\checkmark &   \checkmark  &  \checkmark    &  \checkmark  &  -  &  -  & $986$ &   Transformer & RNNLM $(2 \times 650)$ & 16.0 & 17.1	 \\
\checkmark &   \checkmark  &  \checkmark    &  \checkmark  &  -  &  -  & $986$ &   Transformer & RNNLM $(4 \times 1024)$ & 14.1 & 15.2 \\ 
\checkmark &   -  &  -    &  \checkmark  &  \checkmark  &  \checkmark  & $1460$ &   Transformer & None  & 19.3 & 19.7 \\
\checkmark &   -  &  -    &  \checkmark  &  \checkmark  &  \checkmark  & $1460$ &   Transformer & RNNLM $(2 \times 650)$  & 14.3 & 14.9 \\
\checkmark &   -  &  -    &  \checkmark  &  \checkmark  &  \checkmark  & $1460$ &   Transformer & RNNLM $(4 \times 1024)$ & \textbf{12.3} & \textbf{12.8}\\
\bottomrule
\end{tabular}
\label{table:dsetcompare}
\end{table}

\subsection{Discussion of Results}
The benchmark results are listed in Tab.~\ref{table:dsetcompare}.
First, the effects of using different dataset combinations are inspected.
By using the CommonVoice dataset in addition to Tuda-DE, SWC and M-AILABS,
the test WER decreases to $15.2\%$~WER.
Further replacing SWC and M-AILABS by the custom aligned SWC and Librivox dataset decreased the test set WER down to $12.8\%$.

The second observation is that the language model size and also the achieved perplexity on the text corpus highly influences the WER.
The significant improvement in WER of $2\%$ can be explained by the better ability of the big RNNLM in detection and prediction of German words and grammar forms.
For example, Milde et al.~\cite{Milde2018} described that compounding poses are a challenge for the ASR system;
not recognized compounds resulted in at least two errors, a substitution and an insertion error.
This was also observed in a decoding run without the RNNLM, e.g., \emph{``Tunnel\-ein\-fahrt''} was recognized as \emph{``Tunnel\allowbreak\textbf{\_}ein\allowbreak\textbf{\_}fahrt''}.
By inspection of recognized transcriptions, most of these cases were correctly determined when decoding with language model, even more so with the large RNNLM.

Tab.~\ref{table:dsetcompare} gives us further clues how the benefits to end-to-end ASR scale with the amount of automatically aligned data.
The benchmark results obtained with the small language model improved by absolute $0.1\%$ WER on the Tuda-DE test set,
after addition of the CommonVoice dataset, $319$h of speech data.
The biggest performance improvement of $4.3\%$ WER was obtained with the third selection of corpora with $1460$h of speech data.
Whereas the composition of corpora is slightly different in this selection,
two main factors contributed to this improvement: The increased amount of training data and better utterance alignments using CTC-segmentation.

\section{Conclusion}

End-to-end ASR models require more training data as conventional DNN/HMM ASR systems, as those models grow in terms of depth, parameters and model capacity.
In order to compile a large dataset from yet unlabeled audio recordings, we proposed CTC-segmentation.
This algorithm uses a CTC-based end-to-end neural network to extract utterance segments with exact time-wise alignments.

Evaluation of our method is two-fold:
As evaluated on the hand-labeled dev and test datasets from TEDlium v2, alignments generated by CTC-segmentation were more accurate compared to those obtained from Viterbi- or DTW-based approaches.
In terms of ASR performance, we build on a composition of German speech corpora~\cite{Milde2018} and trained an end-to-end ASR model with CTC-segmented training data;
the best model achieved $12.8\%$ WER on the Tuda-DE test set, an improvement of $1.6\%$ WER absolute in comparison with the conventional hybrid DNN/HMM ASR system.

%
%
%
\bibliographystyle{splncs04}
\bibliography{dt}
\end{document}